# Robust, C-related, Superconducting Nanostructure at the Apex of a W STM Tip

C.G. Ayani[1,2], F. Calleja[1], P. Casado[1], A. Norris[1], J.J. Navarro[1], M. Garnica[1], M. Acebrón[1], D. Granados[1], A.L. Vázquez de Parga[1,2], J.G. Rodrigo[2] and R. Miranda [1,2 *]

[1] Instituto Madrileño de Estudios Avanzados en Nanociencia (IMDEA Nanociencia), Cantoblanco 28049, Madrid, Spain;

[2] Departamento de Física de la Materia Condensada, Instituto Nicolás Cabrera and IFIMAC, Universidad Autónoma de Madrid, Cantoblanco 28049, Madrid, Spain

*corresponding author rodolfo.miranda@imdea.org

**ABSTRACT**

*By pulsing the tunnelling voltage between the W tip of a scanning tunnelling microscope (STM) and a graphene-covered metal surface, a superconducting (SC) nanostructure is formed at the apex of the STM tip. We have characterized the SC properties of the resulting nanotip as a function of temperature and magnetic field, obtaining a transition temperature of 3.3 K and a critical field well above 3T. The SC nanotip is robust, stable, and achieves atomic resolution. A non-SC tip can be easily recovered by controlled voltage pulsing on a clean metal surface. The present result should be taken into account when studying zero-bias features like Kondo resonances, zero-bias-conductance peaks or superconductivity on graphene-based systems by means of STM using tungsten tips.*

The possibility to employ a superconducting (SC) tip in a low temperature STM is extremely useful in a number of experiments that require an increased energy resolution in local spectroscopy [1], the study of tunable Josephson junctions [2-3], the mapping of vortices [4], the determination of absolute spin polarization [5], or the disentanglement of other physical phenomena [6-9]. In most cases, this is achieved by using bulk superconducting materials for the tip (e.g. Nb [1, 4, 6], V [5, 8, 9]) or by indenting a non-SC tip into an SC material (e.g. Pb or Al) [3, 7]. For a number of different experiments, it would be highly desirable to create *in- situ* a superconducting STM tip in a controlled and reversible manner.

We describe a simple method to achieve this enhanced functionality by means of a reproducible functionalization of W tips on graphene covered surfaces, such as gr/Ir(111) or gr/Pb/Ir(111), that produces stable and robust superconducting tip apex. Starting from Ar+ sputtered pristine tungsten tips, we reproducibly create a SC nanostructure at the tip apex by means of controlled voltage pulses on different graphene-covered surfaces. This observation might also be relevant for spectroscopic measurements on graphene-related systems, since the procedure used to clean in-situ the tip of the STM might produce inadvertently the SC nanotip.

The experiments have been carried out in a UHV chamber equipped with a Joule-Thompson STM (JT-STM), a Low Energy Electron Diffractometer (LEED) and facilities for cleaning the samples and STM tips. A single monolayer of graphene was grown by decomposing ethylene on a reactive metallic substrate. Specific metals were intercalated underneath the monolayer of graphene by deposition and gentle annealing [10]. All superconducting gaps were measured under open feedback loop conditions using standard low-frequency ac lock-in detection techniques with a bias modulation of 200 μV peak to peak at 763 Hz. The tips were polycrystalline W or PtIr wires.

The typical sequence of SC nanotip formation starts from freshly Ar$^+$ sputtered polycrystalline W tips, the SC structure being formed when performing controlled voltage pulses (10V, 100 milliseconds, see Suppl. Mat. S1) on graphene-covered surfaces, such as gr/Ir(111) or gr/Pb/Ir(111). The same procedure on graphene-free metal surfaces (e.g. clean Ir(111), Pb/Ir(111), Cu(111)) or using PtIr tips does not result in the formation of a SC nanostructure at the tip apex (see Suppl. Mat. S2).

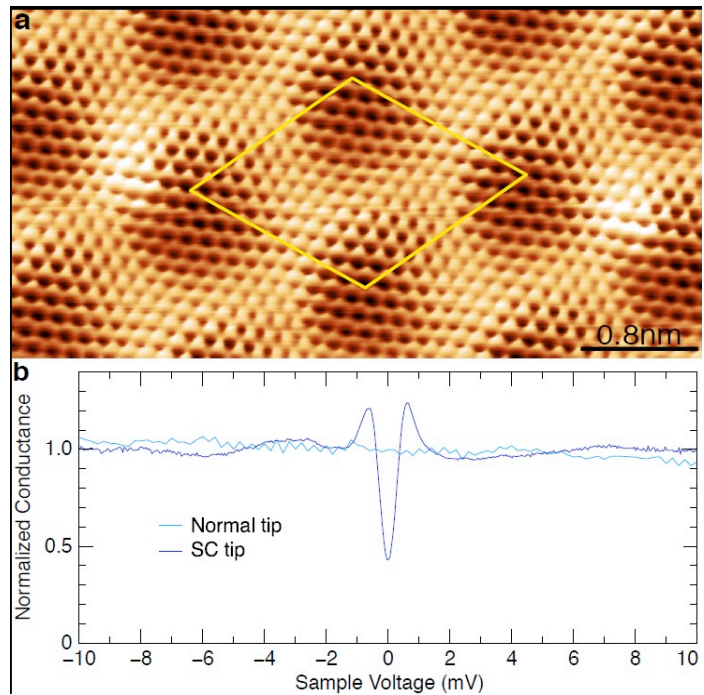

**Fig.1.** *(a) Atomic resolution STM topographic image of graphene/Ir(111) recorded with a superconducting tip prepared by the method described in the main text. The moiré pattern of gr/Ir(111) is highlighted in yellow. Stabilization parameters: V=2 V, I=2 nA; (b) Corresponding differential conductance (dI/dV) of single-particle tunnelling in the STM junction between the normal and modified W tip and graphene/Ir(111) at V=15 mV, I=500 pA. All measurements were performed at T=1.19 K.*

The superconducting nanostructure created at the apex of the tip is mechanically stable (see Suppl. Info. S4) and capable of routine atomic resolution, as shown in Fig. 1a for gr/Ir(111), where in addition to the well-known moiré pattern [11], the hexagons of the atomic lattice of graphene are clearly resolved. Fig. 1b shows the normalized differential conductance of a superconductor/insulator/normal (S/I/N) vacuum tunnel junction consisting of the *in situ* prepared W tip and a graphene/Ir(111) sample kept at T=1.19 K. A well-defined gap develops at the Fermi level. Clear coherence peaks appear at the quasiparticle band edges. Note that the zero-bias conductance does not reach zero, as compared with other bulk superconductors. This makes sense since only the apex of the W tip is SC and not the whole tip, whereby it will introduce a proximity effect on the SC nanotip. Indeed, this pair-breaking phenomenon will have to be taken in consideration when modelling the gap spectra as we will discuss later.

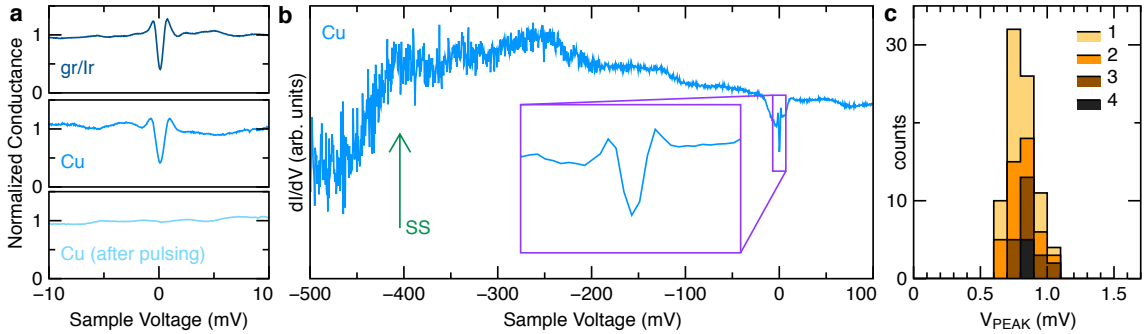

***Fig. 2. (a) Left panel:*** *Differential conductance measured with the SC nanotip on a gr/Ir(111) surface, on a Cu(111) surface and, finally, after pulsing on clean Cu(111). Stabilization parameters V=15 mV, I=2 nA. Modulation voltage 200 µV;* ***(b) Central panel:*** *Large scale STS spectrum measured with the SC nanotip on Cu(111) Note the presence of the superconducting gap at the Fermi level (highlighted in purple) and the onset of the surface state of Cu(111) at around -400 meV (highlighted in green). The noise at high negative bias voltages is due to the small modulation voltage (2 mV) required to resolve the SC gap at the same time. Stabilization parameters V=600 mV, I=3nA;* ***(c) Right panel:*** *Distribution of values of $V_{peak}$ (energy position of the coherence peaks) at 1.1 K for 83 SC nanostructures at the apex of different tips always prepared on gr/Ir(111) and measured on gr/Ir(111), Pb/Ir(111), gr/Pb/Ir(111) and Ir(111) (color codes 1, 2, 3 and 4 respectively)*

In order to demonstrate that the superconductivity resides in the tip and not in the substrate, we have tested these tips, prepared on gr/Ir(111), *in-situ* on a clean Cu(111) surface, finding that the characteristic surface state of Cu(111) *and* the SC gap can be resolved simultaneously in the STS spectra, as demonstrated in the of Fig. 2b. This is a clear proof that the superconductivity resides in the tip.

The tip can be restored into a non-SC state by gently pulsing on graphene-free metal surfaces, as illustrated in Fig. 2a, which shows the gap measured on gr/Ir(111) with a superconducting tip prepared on (a different region of) gr/Ir(111), the gap measured with the same tip on a Cu(111) surface after replacing *in-situ* the gr/Ir(111) sample by the Cu crystal, and, finally, the disappearance of the SC gap after pulsing the same tip on the clean Cu(111) surface.

The results concerning the formation and the value of the SC gap of the tip are robust and easily reproducible. As shown in the histogram of Fig. 2c, data for 83 SC tips measured on different, non-superconducting surfaces (Ir(111), gr/Ir(111), Pb/Ir(111), gr/Pb/Ir(111)) give an average value for the $V_{peak}$ of 0.8 meV with a standard deviation of ± 0.1 meV, i.e. once that the SC apex is formed, its properties are extremely reproducible and independent of whether they are measured in tunnel junctions against different substrates. This large reproducibility is in contrast with the case of Ir tips massively indented on Nb, where the SC gap ranges from 21% to 86% of the bulk Nb value [6].

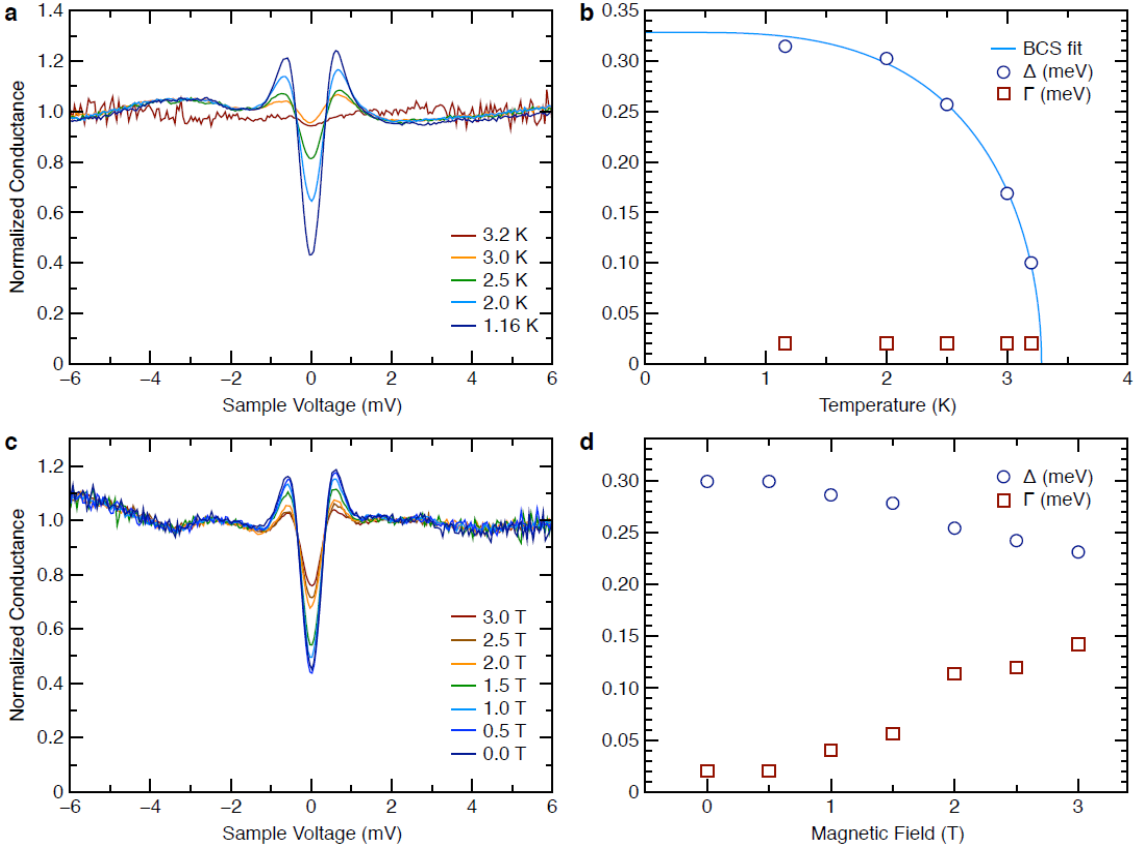

***Fig.3. (a)*** *Temperature dependence of the normalized differential conductance spectra of single-particle tunnelling between a superconducting nanotip and a gr/Ir(111) substrate. The set point was fixed at V=10 mV and I=2.5 nA, i.e. with a tunnelling resistance of 4 MΩ;* ***(b)*** *Superconducting gap (Δ) and pair-breaking parameter in energy units (Γ = ζ Δ) as a function of temperature. The continuous line is the fit to the temperature dependence of the energy gap as given by the BCS equation;* ***(c)*** *Magnetic field dependence of the differential conductance at 1.16 K up to 3 T;* ***(d)*** *Evolution of the parameters of the fit with the external magnetic field.*

The evolution of the gap in the differential conductance with temperature is shown in Fig. 3a. The spectra show the suppression of the quasiparticle coherence peaks with increasing T as well as the increase of the zero bias conductance as the gap is filled.

In order to account for the contribution of the superconducting density of states (DOS) to the measured conductance curves, we have performed a numerical simulation of the spectra. The superconducting DOS is described using Maki formalism [12], where the "standard" BCS superconducting DOS is modified through a pair-breaking parameter, $\zeta$, which accounts for different physical pair-breaking mechanisms, such as those due to proximity effect or the one appearing in a small superconducting specimen in the presence of an external magnetic field.

In this formalism the superconducting DOS, $N_S(E)$, reduces to

$$N_S(E) = Re\left(\frac{u}{\sqrt{u^2 - 1}}\right),$$

With $u$ being defined by the non-linear equation

$$u = \frac{E}{\Delta} + \zeta \ \frac{u}{\sqrt{1 - u^2}} \ .$$

We have used the analytical expression for $u$ presented in [13], where $\Delta$ is the superconducting gap and $\zeta$ the pair-breaking parameter. Only these two free parameters, $\Delta$ and $\zeta$ are used in the simulation of the evolution of the conductance curves vs temperature and magnetic field (see Suppl. Mat. S3).

The resulting superconducting gap as a function of temperature is shown in Fig. 3b (and S3a). The continuous line is the fit to the temperature dependence of the energy gap given by the BCS equation yielding $\Delta(0)$=0.33 meV and $T_c$ = 3.3 K. Note that the fitted value of the gap is smaller than $V_{peak}$. The value of the dimensionless BCS ratio $2\Delta(0)/k_B T_C$ of 2.35 indicates that the superconducting nanotip is in the weak coupling regime ($2\Delta/k_B T_C \leq 3.52$) [13]. A small, fixed, pair-breaking contribution was needed to fit the experimental spectra due to proximity effects of the bulk normal tip on the superconducting nanostructure at the tip apex.

The evolution of the gap at a fixed temperature of 1.16 K with an external magnetic field applied perpendicular to the sample surface (along the tip axis) is shown in Fig. 3c. The spectra show a reduction of the features related to the SC gap under the applied perpendicular magnetic field (maximum of 3 T). The fit shows increasing values of the pair-breaking parameter as field increases, but the SC gap is only slightly reduced even at 3 T. This observation agrees with a progressive increment of pair-breaking effects on the superconducting object at the tip apex as the magnetic field is increased. The 20% reduction of the gap with the external field indicates that the zero temperature perpendicular critical field is certainly much larger than 3 T, as expected for a superconducting condensate that is confined spatially to a very small volume at the apex of the tip [14]. These results are an indication that the superconducting nanotip behaves as a standard s-wave superconductor with an isotropic gap and singlet pairing, following a simple BCS model.

In spite of our efforts to identify by Scanning Electron Microscopy and MicroRaman the nature of the SC nanostructure at the apex, no unequivocal identification has been possible. We speculate that it might be related to the formation of a tungsten-based nanocarbon compound at the tip apex, in agreement with the known fact that superconductivity can be present in tungsten-based amorphous compounds, such as tungsten carbide alloys or dilute solid solutions of C in W [15, 16], as well as C-containing W nanostructures [17]. Unlike the present observation, in these cases, the corresponding critical temperatures (and the measured gaps) depend on the dimension, structure and composition of tungsten carbide, spanning the range from 2.74 to 10 K [15-17]. Another possibility would be the formation of a graphene-based structure at the W tip, since there are some indications that a highly doped graphene layer could be superconducting in a similar temperature range [18].

In conclusion, a stable and robust superconducting nanostructure at the apex of an STM tungsten tip can be reproducibly formed by means of voltage pulses on graphene covered surfaces. The SC gap analysis reveals: $\Delta=0.33$ meV, $H_c > 3$T and $T_c=3.3$ K. This observation is of practical importance for studies of superconductivity in graphene-based systems with STM, as W is a commonly used material for STM tips and voltage pulsing on the surface is a common method to prepare the STM tip.

## Supplementary Material

See supplementary material for the procedure of superconducting tip formation (S1); examples of the appearance of the SC gap on graphene-covered and the lack of it on pristine metal surfaces (S2); the fits of the SC gap dependence with temperature and magnetic field (S3); the stability of the SC gap of a given tip against changes in the tip or different surface structures (S4) and for the SEM characterization of the tip (S5).

## Acknowledgements

This work was partially supported by Ministerio de Economía y Competitividad (MINECO) with Grants FIS2016-75862-P, FIS2017-84330-R, DETECTA ESP2017-86582-C4-3-R and FIS2015-67367-C2-1-P, Comunidad de Madrid (NANOMAGCOST-CM, S2018/NMT-4321, NMAT2D P2018/NMT-4511 and S2018/NMT-4291 TEC2SPACE-CM) and EU-COST Programme NANOCOHYBRI, Action CA16218. M.G. has received financial support through the Postdoctoral Junior Leader Fellowship Programme from "la Caixa" Banking Foundation. IMDEA Nanoscience acknowledges support from "Severo Ochoa" Programme for Centres of Excellence in R&D (MINECO, Grant SEV-2016-0686).

# Robust, C-related, Superconducting Nanostructure at the Apex of a W STM Tip

C.G. Ayani[1,2], F. Calleja[1], P. Casado[1], A. Norris[1], J.J. Navarro[1], M. Garnica[1], M. Acebrón[1], D. Granados[1], A.L. Vázquez de Parga[1,2], J.G. Rodrigo[2] and R. Miranda [1,2]

[1] Instituto Madrileño de Estudios Avanzados en Nanociencia (IMDEA Nanociencia), Cantoblanco 28049, Madrid, Spain;

[2] Departamento de Física de la Materia Condensada, Instituto Nicolás Cabrera and IFIMAC, Universidad Autónoma de Madrid, Cantoblanco 28049, Madrid, Spain

**S1- Procedure of superconducting tip formation**

Fig. S1 shows a representative sequence of voltage pulses carried out with a W tip on a gr/Ir(111) surface at 1 K. The SC gap evolves with the number of pulses until a stable value is reached. Using W tips and graphene-covered substrates the number of voltage pulses required to achieve a SC tip may vary, but on the average, the SC tip develops completely after a sequence of 12 ± 8 voltage pulses. With more than 100 attempts on 2 different graphene-covered substrates, the procedure has a rate of success (in producing the SC apex) larger than 84%. The same procedure using PtIr tips on the very same gr/Ir(111) surface does not result in a SC tip.

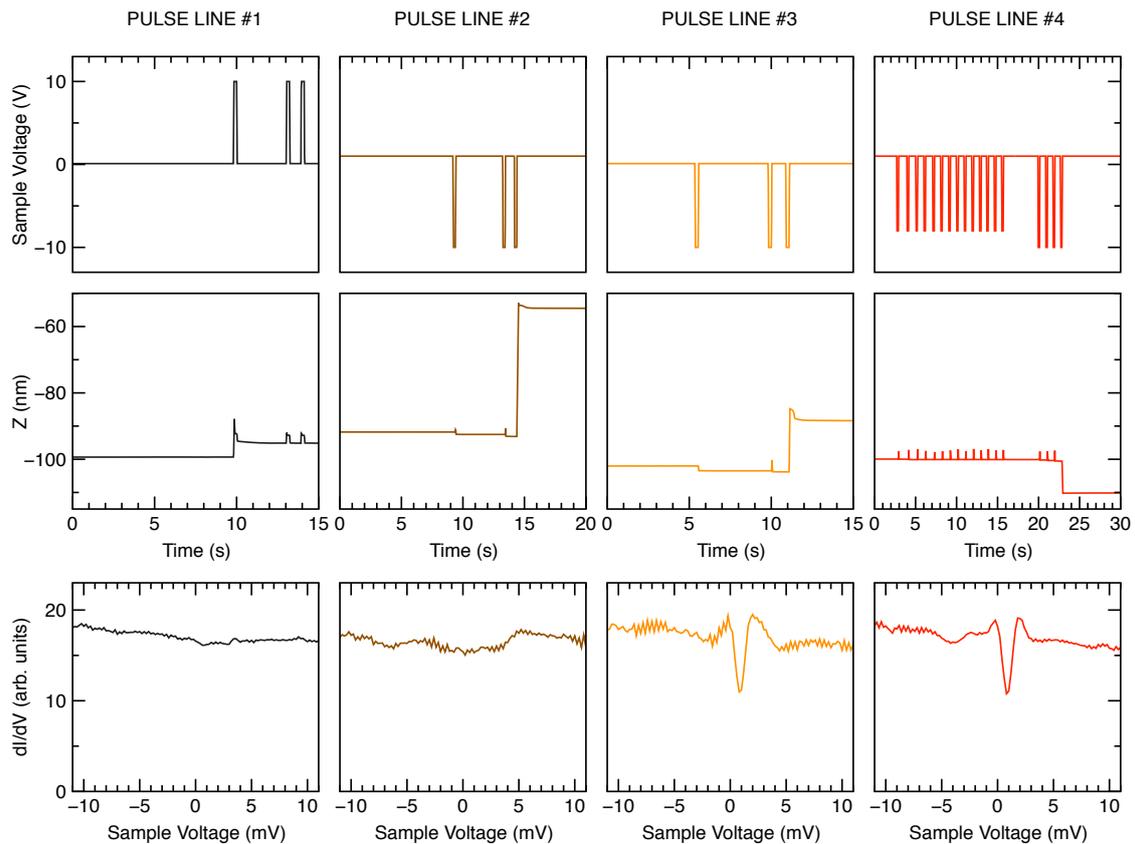

***Fig. S1.*** *Representative sequence of voltage pulses on gr/Ir(111) and the corresponding tip height changes showing the development of the gap due to the superconducting apex formation.*

**S2- Superconducting W-based tip formation on graphene-covered samples**

Two conditions are required for a SC nanotip to be formed after controlled voltage pulses as described in S1: i) a W-based tip in the STM and ii) a graphene-covered surface. This is illustrated below for a W-based tip: On metallic surfaces such as Ir(111) or Pb/Ir(111), the pulses do not produce any SC structure at the apex of the tip. In contrast, on graphene-covered surfaces, e.g. gr/Ir(111) or gr/Pb/Ir(111), a SC nanotip is obtained. On the other hand, performing the same type of sequence with PtIr-based tips does not yield any SC structure, even on the graphene-covered surfaces.

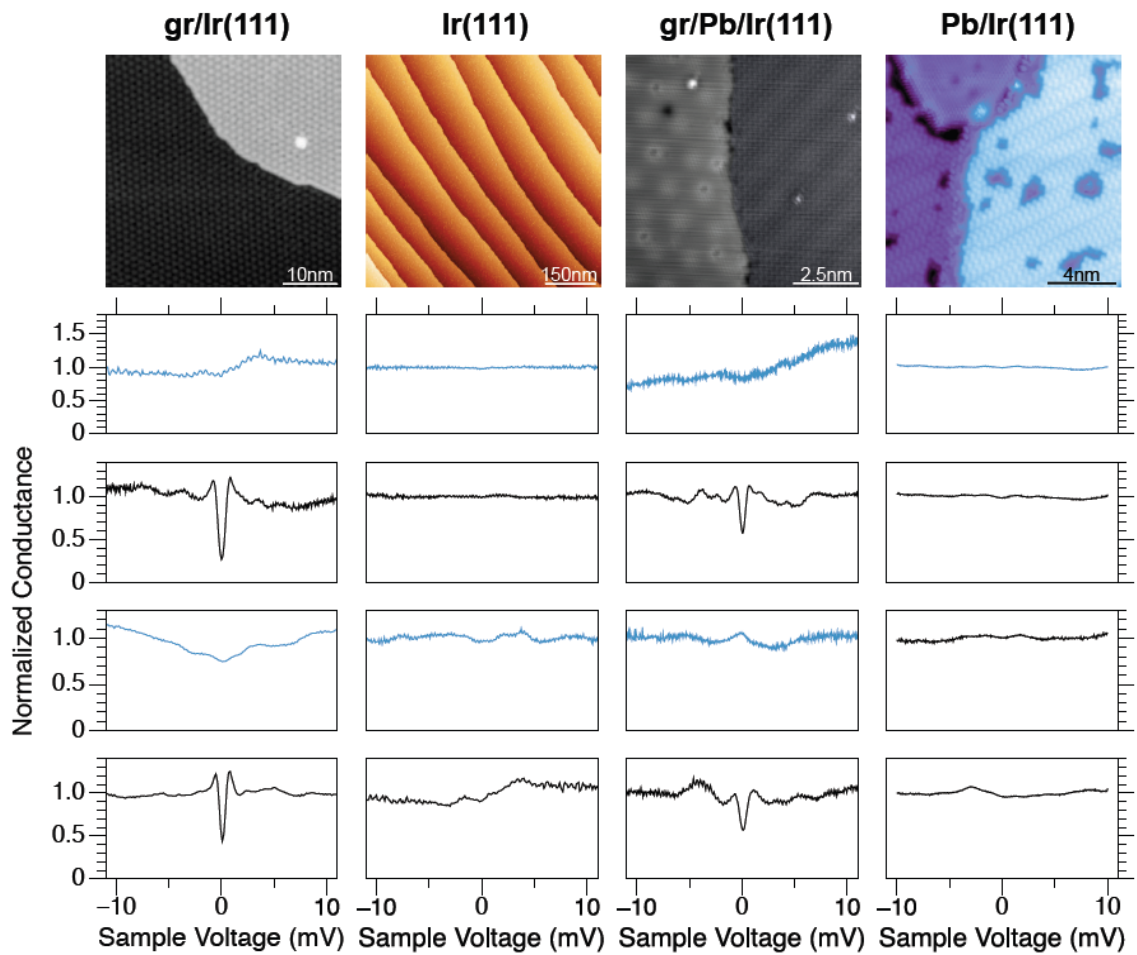

***Fig. S2**: Typical sequence of SC tip formation. Starting from pristine-sputtered tips (blue curves), the SC structure is formed at the apex of the tip after performing voltage pulses (see black curves) only on graphene-covered surfaces at 1 K. These are gr/Ir(111), STM topography image stabilization parameters V=5 V, I= 500 pA, and gr/Pb/Ir(111), STM topography image stabilized at V=-3.1 V and I=3 nA. The same treatment on graphene free surfaces does not result in the SC tip state. These surfaces are Ir(111), STM topography image stabilization parameters V=1 V, I= 700 pA, and Pb/Ir(111), STM topography image stabilized at V=100 mV, I= 3nA. A total of 100 tips were tested, giving 84% success ratio in the formation of the SC structure on the graphene-covered surfaces.*

**S3- Fits of the tunneling spectra**

The differential conductance spectra (normalized to 1 away from the gap) have been fitted to a Density of States (DOS) for the tip that is the combination of the BCS DOS with the Maki expression for the pair breaking due to different physical origins (proximity effects, magnetic field, etc). The DOS of gr/Ir(111) has been supposed to be constant in the relevant energy range. The calculated curves were obtained by convolution of the above mentioned superconducting DOS with the instrumental resolution function of the lock-in amplifier, the corresponding temperature broadening function (Fermi function), and a similar broadening function accounting for electrical noise in our experimental setup (10 $\mu$eV ). The corresponding parameters from the fit have been plotted in Fig. 3 of the main text. Note that this analysis, including different broadening mechanisms, shows that the actual value of the superconducting gap is smaller than the one that could be deduced from the position of the peaks ($V_{peak}$) in the conductance curves.

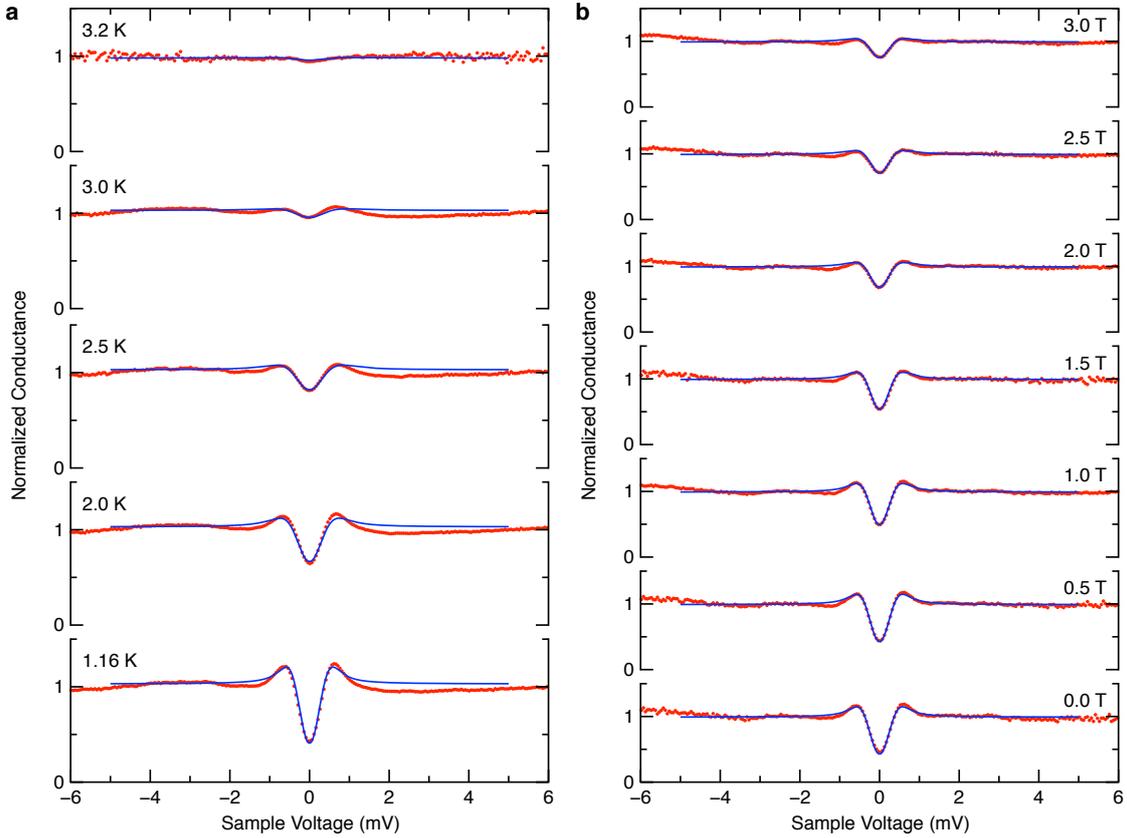

**Fig. S3.** *(a) Temperature dependence of the experimental differential conductance spectra (red curves) of single-particle tunneling between a superconducting nanotip and a gr/Ir(111) substrate. The set point was fixed at 10 mV and i=2.5 nA, i.e. with a tunneling resistance of 4 MΩ. The differential conductance spectra have been normalized to 1 far away from the gap and fitted to the BCS+Maki expression for the DOS (blue curves); (b) Magnetic field dependence of the experimental differential conductance (red) at 1.16 K up to 3 T and calculated conductance (blue).*

**S4- Robust superconducting structure formed at the tip**

The SC nanostructure at the apex of the tip is robust and stable enough to show the same SC gap across graphene regions that have either Pb intercalated underneath graphene or the pristine gr/Ir(111) surface. Even small changes in the tip (Fig.S4c) do not affect the SC gap, nor the high topographic resolution, as demonstrated in Fig. S4 a and d.

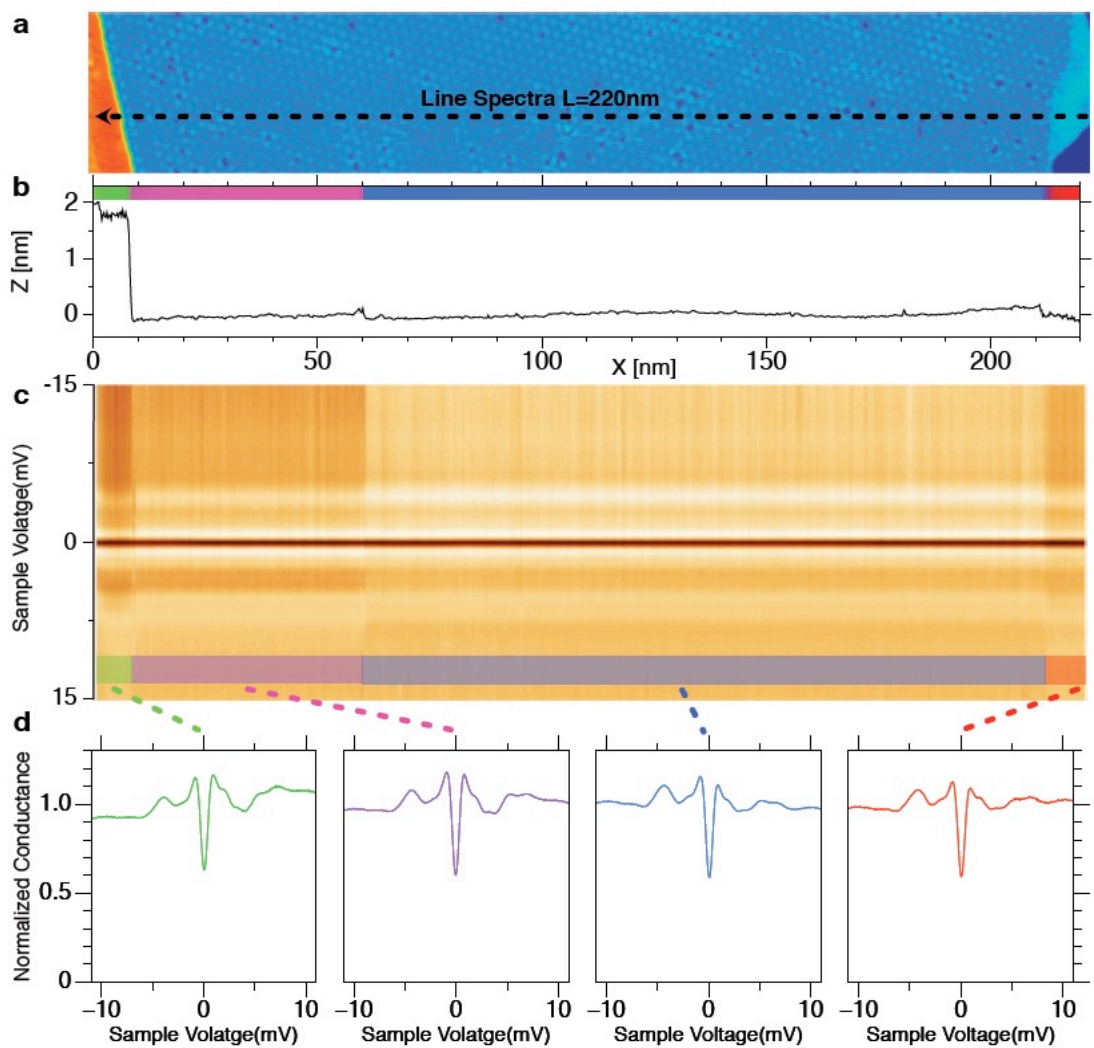

***Fig S4***: *STS line spectra over a large graphene region with different metals underneath. The line is shown on the corresponding STM topographic image (a), with the corresponding height profile (b) and the dI/dV intensity map (c). A total length of 220 nm was explored, starting on a gr/Pb/Ir(111)region at the left, going through a large gr/Ir(111) atomic terrace and ending at the right with another Pb-intercalated graphene region. The gap remains unaltered through the different sample areas (d). Even though some tip changes occur (e.g. between the purple and the blue spectra), the width and height of the gap are maintained. This demonstrates the robustness of the SC structure at the tip apex.*

**S5- Tip structure and preparation**

Our experiments have been performed with homemade and SPECS tungsten tips. Both are electrochemically etched from commercial polycrystalline tungsten wire (99.95% pure) with diameters of 0.3mm and 0.25 respectively. In our case the etching is done in a 5M KOH solution, while SPECS etches the tips in NaOH. A chemical analysis of the wire used in our homemade tips shows the following impurities in ppm: Fe 10, Sn <1, Mg <1, C <10, Cr 4, Cu <1, Pb <1, Al <7, Mo <8, Ni 5, Mn <1, O <25, na <5, Cd <1, Si 4, Co <1, Ca <1, Bi <1, N <10, P <1, Hg <1, K <2.0.

In order to clean the tip apex prior to STM measurements, tips are sputtered by $Ar^+$ with an energy of 2.5 keV for 1-2 hours and sometimes subsequently annealed. This treatment produces non-SC, clean and sharp tips able to perform topographic imaging and STS experiments.

Figure S5 shows a scanning electron microscope (SEM) image of a W tip extracted from the UHV system once the SC gap was obtained. SEM images do not show any clear feature at the apex related to our SC structure.

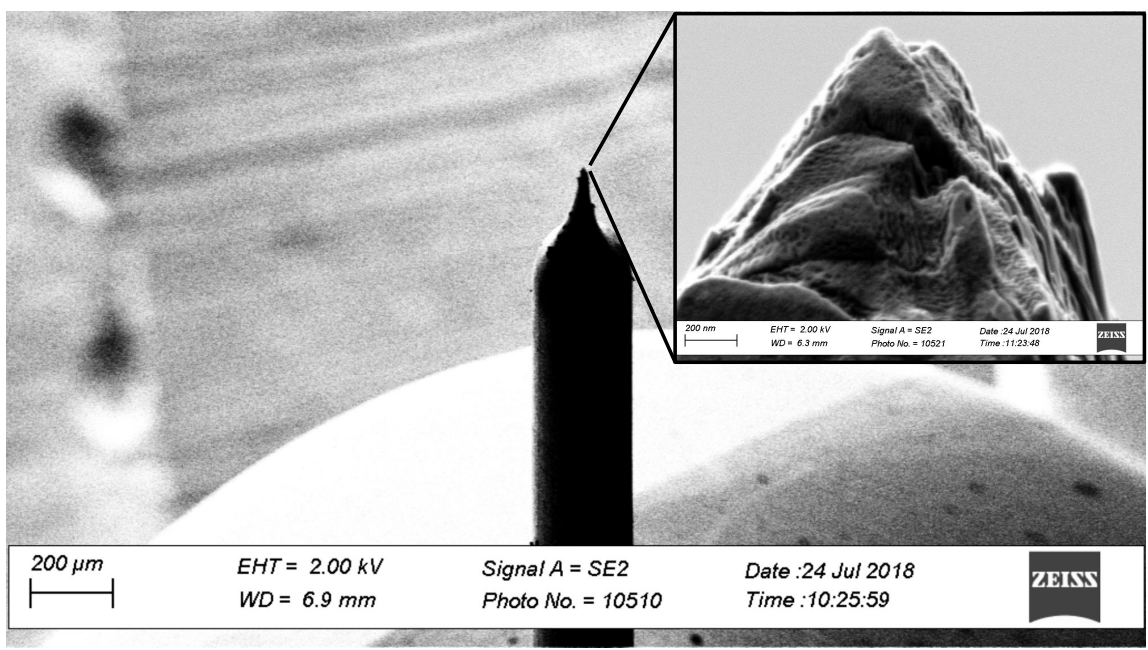

*Fig. S5.* *SEM image of a SPECS tungsten tip after the SC functionalization procedure.*